\documentclass[aps,amsfonts,twocolumn,nofootinbib]{revtex4-1}
\usepackage{epsfig}
\usepackage{graphicx}
\usepackage{slashed}
\usepackage{amsmath}
\usepackage{amsbsy}
\usepackage{bm}
\usepackage{color}
\usepackage{epsfig}
\newcommand{\ee}{\end{equation}}
\newcommand{\bb}{\begin{equation}}
\newcommand{\eqb}{\begin{eqnarray}}
\newcommand{\eqf}{\end{eqnarray}}

\newcommand{\1}{{\'{\i}}}

\def\xvec{\mbox{\boldmath$x$}}
\def\yvec{\mbox{\boldmath$y$}}

\def\1{\'{\i}}

\def\1{\'{\i}}
\begin{document}
\title{$U'(1)$ Neutrino Interaction at Very Low Energies}
\author{J. Gamboa}
\email{jgamboa55@gmail.com}
\affiliation{Departmento de F\1sica, Universidad de Santiago de Chile, Casilla 307, Santiago, Chile}

\begin{abstract} 
 A procedure for studying low energy neutrinos processes  using hadronic physics ideas is proposed. 
We describe how the neutrino-neutrino interactions can be modelled using massive gauge theories. The mass of the 
gauge bosons is  created by spontaneous symmetry breaking and  we show that relationship between mass, currents and potential gauge is through vector dominance, 
i.e. $J_\mu = M^2 A_\mu$. A tiny charge $U(1)$ for neutrinos is turned on for energies around the temperature of the cosmic neutrinos background which allow the  
interaction between neutrinos, above  this temperature this interaction disappears. The magnetic flux is quantized in terms of the  hidden charge.  
We found an approximate relation between the order parameter and neutrino condensate $\langle{\bar \nu} \nu\rangle_{\tiny{0}}$. 
The relation between the neutrino magnetic moment and hidden photons bounds are also discussed.\end{abstract}
\pacs{PACS numbers:}
\date{\today}
\maketitle
%\section{Introduction}

The detection of the cosmic neutrino background (CNB)  is a challenging problem both theoretical and experimental physics because as neutrinos interact very weakly and  practically leave no trace of their existence \cite{libros}.

In recent years many people have argued that the detection of CNB is indirect and its existence could be inferred from the small anisotropies of CMB \cite{refers}. Among some ideas discussed in the literature are the secret \cite{kolb} and electromagnetic neutrino interactions \cite{giunti} which along with possible collective effects of the CNB might produce potentially observables effects. 

Another different possibility is that the low energy neutrinos experience  interaction through its spin \cite{pao}, however for this to happen it is necessary, at least to effective level, that neutrinos have a tiny electric charge (millicharge) that which 
{\it turned on} only at very low energies. The question is, what is the mechanism to turn on the millicharge?. 

For neutrinos there is an upper bound $ e_\nu <10^{-19} $ \cite{Raffelt} which is not ruled out experimentally and, therefore,  could be a window to new physics at the sub-eV scales.

For sub-eV scales an argument in favor of neutrino electromagnetic interaction is not ruled out a priori given the smallness of the $ e_\nu $.

The charge modification $e \to e'$, however, is not free because it requires changing $U(1)$ by $U'(1)$, {\it i.e.} to include hidden photons \cite{holdom}. 

Although the idea of including hidden matter seems sophisticated, this is interesting because the hidden matter  is dark matter. 

In order to realize technically this argument, let's consider the following Lagrangian

\bb 
{\cal L} = {\bar \psi} \left( i{   {{ D}  \hspace{-.6em}  \slash
      \hspace{.15em}}} [A'] - m\right)\psi + {\cal L} (A,A'), \label{hi1}
\ee
where 
\bb 
{\cal L} (A,A') = -\frac{1}{4} F^2 (A) -\frac{1}{4} F^2 (A') + \frac{\gamma}{2} F(A) F(A'). \label{hidd0}
\ee 
and $F_{\mu \nu} = \partial_\mu A_\nu - \partial_\nu A_\mu$. 

Here $A'_\mu$ is a hidden photon that is coupled to a visible one $A_\mu$ by the mixing term $F_{\mu \nu}(A) F^{\mu \nu}(A')$ and $\gamma$ 
must be determined by experiments.

Bounds for $\gamma$  han been extensively discussed  by Jaeckel, Redondo and Ringwald (see {\it e.g.} \cite{jaeckel}).

Diagonalizing in $ A_\mu $ and $ A'_\mu $, (\ref{hi1}) becomes 
 \bb 
{\cal L} = {\bar \psi} \left( i{   {{ D}  \hspace{-.6em}  \slash
      \hspace{.15em}}} [A'] - m\right)\psi + {\cal L} (A, B), \label{hidd0}
\ee
with 
\[ 
{\cal L}(B,A') =  -\frac{1}{4} F^2 (B) -\frac{1}{4} F^2 (A'), 
\]
and 
$B_\mu=A_\mu-\gamma A'_\mu$.  

The covariant derivative in terms of the millicharge $e'$ is  
\[
D_\mu [A'] = \partial_\mu - i e' A_\mu 
\]
with  
\bb 
e'= \frac{1}{\sqrt{1-\gamma^2}}. 
\ee

 The last equation provides the relationship between the visible ($e = 1$) and the $e'$ charges.
 
So even if $\gamma$ is very small (but not zero), the interaction between visible and hidden photons allows neutrinos interact electromagnetically.

In order to determine which parameters will enter into the analysis let us consider the interaction between neutrinos \cite{kolb} 
  
\bb 
{\cal L} =  -\frac{1}{2}\,G_{\tiny{f}}  \,J^2 (x), \label{la22}
\ee
where   $$J^2(x)= \sum_{\alpha, \beta} ({\bar \nu}_\alpha \gamma_\mu \nu_{\beta})^2, $$  
where $G_f$ is a Fermi-like constant and the sum is running over neutrino species.

Linearizing  (\ref{la22}) using the auxiliary vector field $A'_\mu$,  we have
\bb 
{\cal L} =  A'_\mu\, J^\mu +\frac{1}{2G_{\tiny{f}}}A'^2,  \label{laa12}
\ee
and, as first glance, (\ref{laa12}) can be seen as coming from 
\bb 
{\cal L} = -\frac{1}{4} F^2 (A') +\frac{1}{2G_{\tiny{f}}}A'^2 + A'_\mu\, J^\mu.  \label{laa}
\ee

This last \lq \lq massive gauge theory"  is equivalent in the low momentum limit to (\ref{laa12}) if  the kinetic term $F^2$ is neglected, or equivalently,  if the transferred momentum in the process $\nu \nu' \to \nu \nu'$ is negligible. 

After this there are several points to be clarified, namely;  1) If (\ref{laa}) is an approximate description, where does comes from the mass?, 2) What does to neglect the kinetic energy for the hidden photons?, 3) are there new non-perturbative effects?. 

At this level the situation is very similar to the interface between the low energy hadronic physics and  QCD where the concept of approximate symmetry is essential. 

Indeed, in hadronic physics one uses current algebra to derive for example of sum rules \cite{Adler}, here for the same reasons we find results analogous to the vector dominance hypothesis  \cite{yennie} in the electroweak sector, {\it i.e.}  

\bb 
J^a_\mu (x) \simeq \alpha\, A^a_\mu (x), \label{lon22}
\ee 
where $\alpha$ has dimensions of $\mbox{(mass)}^2$. 

The dominance vector hypothesis (\ref{lon22})  and current algebra methods \cite{yennie,CA} are well known established results of hadronic physics 
and they find a very natural translation in superconductivity theory where (\ref{lon22}) corresponds to the London equation where  $\alpha =M$ is the photon mass.

The technical implementation for neutrino physics is as follows; let us start considering the following Lagrangean

\bb 
{\cal L} = {\cal L}_{\varphi A} + {\cal L}_{\nu \phi}, \label{ini}
\ee 
where 
\bb 
{\cal L}_{\varphi A} = \frac{1}{2} |\partial \varphi|^2 + \frac{\mu^2}{2} |\varphi|^2 + \frac{\lambda}{4} (|\varphi|^2)^2 -\frac{1}{4}F^2(A) +{J}_\mu A^\mu, \label{ini1}
\ee 
where 
\bb 
{ J}^\mu = 
\frac{i e'}{2} \left( \varphi^* \partial^\mu \varphi - \partial^\mu \varphi^* \varphi\right) +e'^2 A^\mu |\varphi|^2,    \label{lo23}
\ee
whereas the neutrino-scalar sector is 
\bb  
{\cal L}_{\nu \phi} =    {\bar \nu}\left( i{   {{\tilde D}  \hspace{-.6em}  \slash
      \hspace{.15em}}}+ g_{\tiny Y}  \varphi  - m_\nu \right) \nu  ,  \label{in2}
\ee 
where ${   {{\tilde D}  \hspace{-.6em}  \slash
      \hspace{.15em}}}$ is compact notation containing vector and axial couplings.

      The minimum of the scalar potential is
$$ 
\varphi_0 = \pm \sqrt{\frac{-\mu^2}{\lambda}}. 
$$

If we denote $j^\mu$ the neutrino current, then near of the minimum  of the potential the total current ${\cal J}^\mu = J^\mu + j^\mu$ is  

\bb 
{ {\cal J}}^\mu|_{\varphi= \varphi_0}\simeq j^\mu = M A^\mu ,  \label{higgs1}
\ee
which  is the London equation (\ref{lon22}) with  $M=e' |\varphi_0|$.      

As we will see below this is a clear example that shows how spontaneous symmetry breaking of the algebra and anomalies are related.

At 0-th order in $\varphi$, {\it i.e} in the minimum of the potential,  the effective Lagrangean  becomes 

\bb
{\cal L}  = -\frac{1}{4}F^2(A) + \frac{1}{2}M^2 A^2  +  { { j}}^\mu A_\mu + \cdots, \label{haw}
\ee
which describes an abelian massive gauge field  coupled to an external  fermionic current.  

The dots $\cdots$ are a notation for the free Dirac term ${\bar \nu} (i {   {  \partial\hspace{-.6em}  \slash
      \hspace{.15em}}} - M)\nu$, where $M=m_\nu - g |\varphi_0|$ is a redefinition of the neutrino mass.
      \\

The Hamiltonian analysis of (\ref{haw})  shows that the primary constraint $\chi_1=\pi_0 \approx 0$ leads to the second class one
\bb 
\chi^a_2 =\partial^i\pi^{a}_i + M^2 A^{a}_0 -{ j}^{a}_0 \approx 0,  \label{lig}
\ee 
and so that the Dirac brackets algebra for  $A_0$ and $A_i$ is
\eqb 
\left[ A_i (\xvec), A_j (\yvec) \right] &=&0= \left[ A_0 (\xvec),A_0 (\yvec) \right],  \nonumber
\\ 
\left[ A^{a}_0 (\xvec), A^{b}_i (\yvec) \right]&=&   \frac{1}{M^2}  \partial_i \delta^{(3)} (\xvec-\yvec).   \label{algenon1}
\eqf 

The relationship between $J^\mu$ and $A^ \mu$ is through the equation of motion
\bb 
\partial_\mu F^{\mu \nu} (A) = -M^2 A^{\nu} +j^\nu.   \label{equa}
\ee 

In order to take the low energy limit we proceed as follows;  rescaling  $A'_\mu= M^2A_\mu$, (\ref{equa}) becomes 

\bb \frac{1}{M^2} \partial_\mu F^{\mu \nu} (A')\approx 0 \, \Rightarrow - A^{'\nu} + {j}^\nu =0,  \label{nu0}
\ee 
 and therefore  we obtain the London equation

 \bb 
 j^\mu = M^2 A^\mu.  \label{00}
 \ee
 \\ 
\indent In this approximation and using (\ref{algenon1}) and (\ref {00}) we obtain the following current algebra
\eqb
\left[ j_i (\xvec), j_j (\yvec) \right]  &=&0=\left[ j_0 (\xvec),j_0 (\yvec) \right],  \nonumber
\\ 
\left[ j_0(\xvec), j_i (\yvec) \right]&=&   c\, \partial_i \delta^{(3)} (\xvec-\yvec) ,  \label{algebra}
\eqf 
with the  \lq central charge\rq  $\,\,c=M^2$.  
\\ 

Therefore the Lagrangean  (\ref{haw}) can be written 
\bb 
{\cal L} =  \frac{M^2}{2} A^2 + j_\mu A^\mu + \cdots, \label{domi1}
\ee  
or equivalently as the  four-fermion theory (\ref{la22}) identifying $G_{\tiny{f}}$ with $G=1/M^2$. 

We should also note that the current in this Lagrangian also as a gauge field that transforms as 
$$
j_\mu \rightarrow j_\mu + \partial_\mu \Lambda, 
$$
which is a consequence of the London equation.

Now consider a small displacement around the minimum of the potential $$\varphi = \varphi_0 + \phi, \,\,\,\,\,\,\,\,\,\,\,\, |\varphi_0| << |\phi|$$  
then the total current has the following form 

\bb
 { {\cal J}}^\mu = \frac{i e'}{2}\left( \phi^* \partial^\mu \phi - \phi \partial^\mu \phi^* \right) + e'^2 |\phi|^2A^\mu  + M^2 A^\mu. \nonumber
\ee

Making the change of variables  \cite{NO}  $\phi = |\phi| \, e^{i \chi}$, last equation becomes  

\bb
{\cal J}^\mu = e'|\phi|^2 \partial^\mu \chi  + \left( M^2 +  e'^2 |\phi|^2\right) A^\mu. \label{corre2}
\ee

As  $|\varphi_0| << |\phi|$ and $|\varphi_0| \propto M$ --the low energy limit--  we have 
\bb
{\cal J}^\mu \approx e' |\phi|^2 \partial^\mu \chi  + e'^2 |\phi|^2 A^\mu. \label{corre22} 
\ee

Integrating in the path $C$ in the region without currents, the magnetic flux becomes

\eqb
\Phi &=& \oint_C dx_\mu A^\mu = -\frac{1}{e} \oint_C dx_\mu \partial^\mu \chi = -2 \frac{\pi }{e'}n   \nonumber 
\\
&=& -2\pi n \sqrt{1-\gamma^2}  \,\,\,\,\,\,\,\,\,\,\,\,\,\, n \in Z.
\eqf
 
 Therefore the flux  is quantized in  unit of $\sqrt{1-\gamma^2}$.
\\ 
The flux quantization is a geometrical  implication of the ${\cal J}^\mu$ structure and the London equation. However,  the vortex configuration for the scalar scalar field  is more difficult to prove.  

However even though we cannot find explicit classical solutions one can,  by using heuristic arguments,  find a relation between the order parameter and the neutrino condensate. 
 
Indeed, if  $\langle{\bar \nu} \nu\rangle_{\tiny{0}}$ denotes the neutrino condensate in the bottom of the potential,  then the total  energy  of the scalar-neutrino system is proportional to the potential energy. 

In other words,  the effective lagrangian must have the form
\[
{\cal L}_{eff} = \frac{\mu^2}{2} \phi^2 + g_{\tiny{Y}}  \langle{\bar \nu} \nu\rangle_{\tiny{0}} \phi+ \cdots 
\] 
where $g_{\tiny{Y}}$ a Yukawa coupling constant \cite{bur} and $\cdots$ are negligible contributions in the bottom of the potential. 
\\ 

The classical solutions for $\phi$ is 
\bb 
\langle{\bar \nu} \nu\rangle_{\tiny{0}} =  \frac{\lambda |\varphi_0|^2}{2 g_{\tiny{Y}}} \phi
\ee 
which is similar to the relation between Cooper pairs and the order parameter of superconductivity theory.  
\\

In the same spirit of the photon-hadron interaction, the gauge boson can also carry color quantum numbers and therefore, one expects that the non-Abelian version of the previous analysis also work.

The non-abelian Gauss law now is 

\bb 
\chi^a_2 =\partial^i\pi^{a}_i - g f^{abc} \pi^b_i A^{ci} + M^2 A^{a}_0 -{\cal J}^{a}_0 \approx 0,  
\ee 
so that the Dirac brackets algebra for  $A^{a}_0$ and $A^{a}_i$ becomes 
\begin{widetext}
\eqb 
\left[ A^{a}_i (\xvec), A^{b}_j (\yvec) \right] &=&0, \nonumber
\\ 
\left[ A^{a}_0 (\xvec),A^{b}_0 (\yvec) \right] &=& \frac{g}{M^2} f^{abc} A^c_0 (\xvec)\delta^{(3)} (\xvec -\yvec), \label{algeno0}
\\ 
\left[ A^{a}_0 (\xvec), A^{b}_i (\yvec) \right]&=&   \frac{1}{M^2} \left(\delta^{ab} \partial_i \delta^{(3)} (\xvec-\yvec) +g f^{abc} A^c_i (\xvec)  \delta^{(3)} (\xvec -\yvec)\right).   \nonumber
\eqf 
\end{widetext} 

In this  non-abelian case the low energy limit is more subtle because the London equation contains quadratic powers in A. 

Indeed, as the strength tensor is 

\bb  
F_{\mu \nu} = \partial_\mu A_\nu -  \partial_\nu A_\mu + g [A_\mu, A_\nu],  \label{la}
\ee 
the equation of motion implies 
\bb 
{\cal J}^{a \mu} =  M^2 A^{a \mu} +  D^\mu (A)  F^a_{\mu \nu}. 
\ee

Rescaling again the Yang-Mills potential as $A'^{\mu}=M^2 A^\mu$ and taking the large-M limit  we have 
\begin{widetext}
 \eqb
J^a_\mu &=& A^{'a}_ \mu + \left( \partial_\mu + i \frac{g}{M^2} A^{'}_ \mu \right) \left( \frac{1}{M^2} (\partial_\mu A^{'}_\nu - \partial_\nu A^{'}_\mu) + 
\frac{g}{M^4} [A^{'}_\mu ,A^{'}_\nu)  \right] \nonumber 
\\ 
&\simeq& A'^{a}_ \mu
\eqf 
\end{widetext}
which is again a London equation, {\it i.e.} (\ref{lon22}).

In this limit the current algebra is  

\begin{widetext}
\begin{eqnarray}
\left[ J^a_i (\xvec), J^b_j (\yvec) \right]  &=&0, \nonumber 
\\ 
\left[ J^a_0 (\xvec),J^b_0 (\yvec) \right] &=& g  f^{abc} J^{c}_0 (\xvec)  \delta^{(3)} (\xvec-\yvec),  \label{algenon2}
\\ 
\left[ J^a_0(\xvec), J^b_i (\yvec) \right]&=&     M^2\,\delta^{ab} \partial_i \delta^{(3)} (\xvec-\yvec) +  g f^{abc} J^c_i (\xvec)  \delta^{(3)} (\xvec-\yvec).  \nonumber
\end{eqnarray}
\end{widetext}

We note that the central charges in (\ref{algebra}) and (\ref{algenon2}) are exact ones and  both are solutions of the integral equation 
\[ 
\int_{M^2}^\infty \frac{dm^2}{m^2} \rho (m^2) = M^2,
\]
where $\rho (m^2)$ is the spectral function.  

Thus,  we have the Lagrangean 
\bb  
{\cal L}_C  \simeq -\frac{1}{2 }M^2 A^{a\mu}A_{a\mu} + {{\cal J}^a}^\mu A^a_\mu + .... \label{domi2} 
\ee
which is an explicit realizations of the vector dominance \cite{lee}.
 
The abelianization used in the Yang-Mills case is reminiscent of the 't Hooft abelian projection  \cite{hooft,nambu,hooft,poly}.

In conclusion, in this paper we have proposed how to implement the electromagnetic interactions of neutrinos and hidden photons. 
This interaction is turned when the neutrino energy is around of the CNB temperature and over this energy the interaction becomes is negligible. 
Despite the neutrino and hidden photons interaction cannot be ruled out, this must be consistent with the neutrino magnetic moment, {\it i.e.}   
\[
\mu' -1 =\frac{\gamma^2}{2} + {\cal O} (\gamma^4), 
\] 
which is compatible with the smallness bound for the neutrino.

I am grateful to J. L\'opez-Sarri\'on, F. M\'endez and M. Loewe for discussions.  I would to thank also to C. Garc\1a-Canal and S. L. Adler by the E-mail correspondence. 
This work was supported by a grants from FONDECYT-Chile 11300020.

\end{document}